\documentclass[twocolumn,showpacs,preprintnumbers,amsmath,amssymb]{revtex4}
\usepackage[dvips]{graphicx} 
\usepackage[english]{babel}

\begin{document}

\preprint{IFF-RCA-10-11}
\title{Decoherence in an accelerated universe}
\author{S. Robles-P\'{e}rez, A. Alonso-Serrano and P. F. Gonz\'alez-D\'{\i}az}
\affiliation{Colina de los Chopos, Centro de F\'{\i}sica ``Miguel Catal\'{a}n'',
Instituto de Matem\'{a}ticas y F\'{\i}sica Fundamental,
Consejo Superior de Investigaciones Cient\'{\i}ficas, Serrano 121, 28006
Madrid (SPAIN) and \\ Estaci\'{o}n Ecol\'{o}gica de Biocosmolog\'{\i}a, Pedro de Alvarado, 14, 06411-Medell\'{\i}n, (SPAIN).}
\date{\today}

\begin{abstract}
In this paper we study the decoherence processes of the
semiclassical branches of an accelerated universe due to their
interaction with a scalar field with given mass. We use a third
quantization formalism to analyze the decoherence between two
branches of a parent universe caused by their interaction with the
vaccum fluctuations of the space-time, and with other parent
unverses in a multiverse scenario.
\end{abstract}

\maketitle

\section{Introduction}

Decoherence plays a fundamental role in quantum mechanics and cosmology. It effectively collapses the wave function from a superposition of states to the observed component \cite{Kiefer2007}. In particular, decoherence is responsible for the collapse of the wave function and becomes the ultimate reason for the appearance of a classical universe in quantum cosmology \cite{Kiefer2007, Joos2003}.

In a decoherence process we have first to identify the system under study and its environment, i.e. we have to distinguish relevant from irrelevant variables \cite{Kiefer2007}. This sort of choice is usually based on the particular features of the experiment, though it can become somehow arbitrary. Decoherence clearly depends on that choice and it is therefore related to the separability of the whole quantum system into the system under study and its environment. Thus, different subsystems can effectively be considered  to be the environment of a quantum system.

Moreover, decoherence and dissipative processes between a system and its environment make the state of the system  evolve to a state of higher entropy \cite{Kiefer2007}. In quantum cosmology, such a entropy increasing provides us with an arrow of time and it is  thus the responsible for the irreversibility in the universe. Strictly speaking, time can only be considered  when a decoherence process has taken place and the semiclassical branches of the universe have emerged.

The interaction between an homogeneous and isotropic universe and the density fluctuations and gravitational waves has been studied in Refs. \cite{Zeh1986, Halliwell1989, Kiefer1992,Coleman1988, Coleman1988b, PFGD1992b}. In this case, the relevant variables are the scale factor and the homogeneous degrees of freedom of the scalar field. The non-observable degrees of freedom, which are traced out from the state of the universe, have however observable effects on the properties of the semiclassical branch of the universe.

Among these effects, it can be pointed out: i) the decoherence of different branches of the universe \cite{Halliwell1989, Kiefer1992}, the shift of the coupling constants and the reduction of the value of the cosmological constant \cite{Coleman1988, Coleman1988b}, and the modification of the coherence properties of the fields that propagate in the space-time \cite{Coleman1988, PFGD1992b}.

In this paper, we review the effects that  decoherence processes can produce on the state of an homogeneous and isotropic branch of an accelerating universe. In Sec. II, we apply the formalism developed in Refs.  \cite{Halliwell1989, Kiefer1992}  to analyze the decoherence between the expanding and contracting branches of the universe due to the interaction with a scalar field, for  quintessence-dominated, vacuum-dominated and phantom-dominated universes. The third quantization formalism is used in Sec. III to study the  interaction  between parent universes and between a parent universe and a plasma of baby universes which represent, in a first approximation, the quantum fluctuations of the space-time of the parent universe, following the parallel quantum optics developments. In Sec. IV,  we shall draw some conclusions.

\section{Decoherence of the branches of an accelerating universe}

In Refs. \cite{Halliwell1989, Kiefer1992, Kiefer2007}, the inhomogeneous modes of a scalar field are taken as the irrelevant variables in order to obtain a reduced density matrix that represented the quantum state of the homogeneous universe. The inhomogeneous modes were coupled  to the metric of a spatially closed space-time \cite{Halliwell1985}, which is assumed to be in the semiclassical regime. The result is that the expanding and contracting branches of the universe are quickly decoupled for large values of the scale factor. Thus, the universe is either in an expanding or in a contracting state but not in a quantum superposition of both.

In a flat universe, the formalism used in Ref. \cite{Halliwell1989} cannot be directly applied because the inhomogeneous modes of the massless scalar field are coupled to a zero curvature term of the metric  \cite{Halliwell1985}. However, we can consider a scalar field with a mass term which is minimally coupled to the curvature scalar. Except for this feature, in order to analyze the decoherence of the branches of the accelerated universe, we can follow the same procedure used in Ref. \cite{Halliwell1989}.

Let us consider therefore a flat homogeneous and isotropic universe which is dominated by a perfect fluid with equation of state $p = w \rho$, where $p$ and $\rho$ are the pressure and the energy density of the fluid, respectively, and $w$ is a constant parameter. Let us also consider a scalar field $\varphi$ with mass, $m$. The Wheeler-De Witt equation can be written, with the usual choice of the factor ordering \cite{Kiefer2007}, as
\begin{equation}\label{WDW1}
 \left( a^2 \partial^2_{a a} + a \partial_a + \frac{\Omega_0^2}{\hbar^2} a^{2 q}  - \partial^2_{\varphi \varphi} + \frac{m^2}{\hbar^2} a^6 \varphi^2 \right) \phi(a, \varphi) = 0 ,
\end{equation}
where $a$ is the scale factor, $\phi(a, \varphi)$ is the wave function of the universe, $\Omega_0^2$ is a constant which is proportional to the energy density of the universe at a given boundary hypersurface at $a_0$ \cite{PFGD2009}, and, $q = \frac{3}{2}(1-w)$. The solutions of the gravitational part of Eq. (\ref{WDW1}) are given in terms of Bessel functions \cite{PFGD2007, RP2010},
\begin{equation}\label{phi0}
\phi_0(a) \propto \mathcal{H}_0^{(1,2)}(\tilde{\Omega}_0 a^q) ,
\end{equation}
with $\tilde{\Omega}_0\equiv \frac{\Omega_0}{q \hbar}$, and, $\mathcal{H}_n^{(1)}$ and $\mathcal{H}_n^{(2)}$ are the Hankel function of first and second kind of order $n$. The boundary condition that has been used in Eq. (\ref{phi0}) is the tunneling boundary condition \cite{Vilenkin1986}. In the asymptotic limit of large values of the scale factor \cite{Abramovitz1972}, we have
\begin{equation}\label{eq238}
\mathcal{H}_0^{(1,2)}(\tilde{\Omega}_0 a^q) \sim a^{-\frac{q}{2}} e^{\pm i (\tilde{\Omega}_0 a^q -\frac{\pi}{4})} ,
\end{equation}
where the $+$ and $-$ signs correspond to the Hankel function of the first and second kind, respectively. The solutions given by Eq. (\ref{eq238}) describe the expanding and contracting branches of the semiclassical universe, as it can be checked  by noting that the momentum operator is given by, $\hat{p}_a \phi(a) \equiv -i \hbar \partial_a \phi(a) \propto \pm \frac{\partial S_0}{\partial a}$, with $S_0 = \tilde{\Omega}_0 a^q$  the classical action, and the semiclassical momentum becomes, $p_a^{sc} = - a \dot{a}$. Then, $a \dot{a} \propto \mp \frac{\partial S_0}{\partial a}$. Thus,  the Hankel function of the second kind corresponds to the expanding branch of the universe and the Hankel function of the first kind describes its contracting branch.

In the semiclassical regime, the wave function   of the universe can be written as,
\begin{equation}
\phi(a, \varphi) = C(a) e^{\pm i S_0(a)} \chi(a, \varphi) ,
\end{equation}
where $C(a) = a^{-\frac{q}{2}}$. The function $\chi(a, \varphi)$ satisfies the following Schr\"{o}dinger equation \cite{Halliwell1989, Kiefer1992},
\begin{equation}\label{eq532}
\frac{1}{2 a^3} \left( -\frac{\partial^2 }{\partial \varphi^2} + m^2 a^6 \right) \chi = i \frac{\partial \chi}{\partial t} ,
\end{equation}
where the time variable $t$ is defined in terms of the scale factor through the classical equation, $a \dot{a} = \pm \Omega_0 a^{q-1}$. Following the above references, we  shall look for Gaussian solutions
\begin{equation}\label{eq533}
\chi = A(a) e^{- B(a) \varphi^2} .
\end{equation}
Inserting Eq. (\ref{eq533}) into Eq. (\ref{eq532}), we obtain a differential equation for the coefficients  $A(a)$ and $B(a)$ that can be solved with the normalization condition,  $ \int d\varphi \chi \chi^* = 1$. It is obtained
\begin{eqnarray}
A &=& \pi^{-\frac{1}{4}} (B + B^*)^{\frac{1}{4}} e^{i \alpha(t)} , \\ B &=& \pm i \frac{\Omega_0}{2} a^{q+1} \frac{\dot{x}}{x}  ,
\end{eqnarray}
where, $\dot{x}\equiv \frac{\partial x}{\partial a}$, and
\begin{equation}\label{eq538}
a^2 \ddot{x} + (1+q) a \dot{x} - \frac{m^2}{\Omega_0^2} a^{2(3-q)} x = 0 .
\end{equation}
For a universe dominated by a cosmological constant ($w=-1$ and $q=3$), the solutions to Eq.  (\ref{eq538}) can be written as,
\begin{equation}
x = a^{-\frac{3}{2}} \left( c_1 a^{k_1} + c_2 a^{-k_1} \right) ,
\end{equation}
with, $k_1 = \frac{3}{2}\sqrt{1+\frac{4 m^2}{9 \Omega_0^2}} \approx \frac{3}{2}$. Choosing appropriate constants $c_1$ and $c_2$  to fulfill the condition, $B + B^* >0$, the value of the fucntion $B(a)$ can be approximated, in the semiclassical regime, as
\begin{equation}
B_0^\pm \approx \frac{\Omega_0}{2} \left( 3 \pm i \frac{m^2}{3 \Omega_0^2} a^3 \right) ,
\end{equation}
where the positive and negative sign correspond to the solution of the function $\chi$ for the expanding and the contracting branches of the universe, respectively.

In the quintessence regime, for which  $-\frac{1}{3}> w > -1$, the solutions of Eq. (\ref{eq538})  can be written in terms of the modified Bessel functions of order $\nu$, $\mathcal{K}_\nu$ and $\mathcal{I}_\nu$. In the semiclassical regime, it reads
\begin{equation}\label{eq5312}
( B^-_Q)^* = B^+_Q \approx \frac{\Omega_0 }{2} \left( i \frac{m}{\Omega_0} + \pi (3+2|\beta|) a^{-\beta} e^{- 2\lambda a^\beta} \right) ,
\end{equation}
where, $q=\frac{3}{2}(1-w) \equiv 3 - \beta$ (with, $1>\beta>0$), and $\lambda = \frac{m}{\Omega_0 |\beta|}$.

In the phantom regime, for which  $w<-1$ and $\beta <0$, the functions $B_F^\pm$ can be approximated in the semiclassical regime by,
\begin{equation}\label{eq5314}
( B^-_F)^* = B^+_F \approx \frac{\Omega_0 }{2} \left( - 2i q a^q + c_0\right) ,
\end{equation}
with $q = 3 + |\beta|$, and  $c_0$  a positive constant.

Inserting these values of the functions $A(a)$ and $B(a)$ in Eq. (\ref{eq533}), the reduced density matrix becomes
\begin{equation}
\rho_r(a, a') = \int_{-\infty}^\infty d\varphi \chi^*(a, \varphi) \chi(a', \varphi) ,
\end{equation}
which is given, except for irrelevant phases, by \cite{Halliwell1989}
\begin{equation}\label{eq5316}
\rho_r(a, a') \propto  \left[ \frac{( B(a) + B^*(a)) (B(a') + B^*(a')) }{( B^*(a) + B(a') )^2}  \right]^{\frac{1}{4}} .
\end{equation}
For the case $w = -1$, Eq. (\ref{eq5316}) can be approximated in the semiclassical limit as,
\begin{equation}\label{eq5317}
\rho_r(a,a') \approx \frac{1}{ \sqrt{1 \mp \frac{i m^2}{18 \Omega_0^2} (a^3 - a'^3)   } } .
\end{equation}
The diagonal values of the reduced density matrix, for which
$a\approx a'$,  become nearly unity. However, far from the
diagonal elements, for which $a\gg a'$, the reduced density matrix
asymptotically vanishes, $|\rho_r| \sim a^{-\frac{3}{2}}$. That
means that the decoherence process between the branches with
different values of the scale factor  is rather effective for
large values of the scale factor, such as it should be expected.

In the quintessence regime,$-\frac{1}{3}\geq \omega \geq -1$ the
decoherence process turns out to be even more effective. The
reduced density matrix (\ref{eq5316}) can then be approximated as,
\begin{widetext}
\begin{equation}\label{eq5318}
\rho_r(a, a') \approx \sqrt{c} \left[  \frac{ (a a')^{3-\beta} e^{- 2 \lambda (a^\beta - a'^\beta)}    }{ \left( \mp i \frac{m}{2} (a^3 - a'^3) + \frac{c}{2} ( a^{3-\beta} e^{-2\lambda a^\beta} + a'^{3-\beta} e^{-2\lambda a'^\beta} )   \right)^2      }   \right]^{\frac{1}{4}} .
\end{equation}
\end{widetext}
For the diagonal values we have $\rho_r(a, a) \approx 1$, and for
the off-diagonal values,
\begin{equation}
|\rho_r (a, a')| \sim \left( \frac{a'}{a}\right)^{\frac{q}{2}}
a^{-\beta} e^{- \lambda a^\beta} \rightarrow 0 \;\; (a\gg a' >1).
\end{equation}

Finally, for the phantom regime, $w<-1$ and $\beta <0$. For large
values of the scale factor in that regime, but still before
reaching the achronal region around the big rip singularity, where the
semiclassical approximation is no longer valid
\cite{Dabrowski2006},
\begin{equation}\label{eq5320}
\rho_r(a, a') \approx \frac{1}{\sqrt{1 \pm i \frac{q}{c_0} (a^q - a'^q) }   } ,
\end{equation}
and, $|\rho_r| \sim a^{-\frac{q}{2}}$, for $a\gg a'$. When the
universe approaches the big rip singularity,  the state of the
universe is  given by a quantum superposition of states
\cite{Dabrowski2006}, concordant with the expected quantum nature
of the universe in such a region \cite{Dabrowski2006, Nojiri2004}.

It can be concluded that, both in a contracting and an expanding
branch of an accelerated  universe, the decoherence process
between the scale factor and a scalar field is effective enough to
remove the quantum interference between the different
semiclassical  branches that correspond to different values of the
scale factor.

The same decoherence process turns out to be also effective to
eliminate the interference between the contracting and expanding
branches. If the state of the universe is given by a quantum
superposition of the states that correspond to those branches,
i.e.
\begin{equation}
\phi(a) \approx C(a) e^{- i S_c(a)} \chi^+(a, \varphi) + C^*(a) e^{ i S_c(a)} \chi^-(a, \varphi) ,
\end{equation}
with, $\chi^+ = (\chi^-)^*$, then, the reduced density matrix will
show four terms \cite{Halliwell1989}: the terms $\rho_{11}$ and
$\rho_{22}$, which describe the quantum state of the expanding and
contracting branches, respectively, are given by the expressions
given above (Eqs. (\ref{eq5317}, \ref{eq5318},\ref{eq5320})). The
crossed terms, which correspond to the interference between the
branches, are given by
\begin{equation}
(\rho_{21})^* = \rho_{12} = \int_{-\infty}^\infty d\varphi \chi^+(a, \varphi) (\chi^-(a', \varphi))^* .
\end{equation}
They turn out to be \cite{Halliwell1989},
\begin{equation}
\rho_{12}(a, a') \propto \left[ \frac{( B^+(a) + \bar{B}^+(a) ) (B^-(a') + \bar{B}^-(a') ) }{( B^+(a) + \bar{B}^-(a') )^2}  \right]^{\frac{1}{4}} ,
\end{equation}
where, $\bar{B} \equiv B^*$. In that case, even for similar values
of the scales factors, $a \approx a'$, the elements of the reduced
density matrix, $\rho_{12}$ and $\rho_{21}$, asymptotically vanish
when the scale factor grows along the semiclassical regime. For
instance, for a  phantom dominated universe, $\rho_{12}$ turns out
to be
\begin{equation}
\rho_{12} \approx \frac{1}{\sqrt{1- i \frac{q}{c_0} (a^q + a'^q) } } ,
\end{equation}
and the diagonal values,
\begin{equation}
|\rho_{12}(a, a)| \approx a^{-\frac{q}{2}} .
\end{equation}
It means that the expanding and contracting branches of a phantom universe, which correspond to the regions far before and after the big rip singularity, decouple from each other  in the semiclassical regime.

Therefore, the decoherence process between the scale factor and a
scalar field with mass is seen to be effective enough to remove
the interference terms between the different semiclassical
branches of an accelerated universe. It eliminates both the
interference terms between the expanding and  contracting
branches, and those between different branches that correspond to
different values of the scale factor in the same expanding or
contracting region of the universe.

In the case of a universe dominated by phantom energy, the big rip
singularity makes it impossible a semiclassical  description of
the universe in the neighborhood  of the singularity. There,  the
state of the universe is given by a quantum superposition of
states \cite{Dabrowski2006} and the quantum effects are
predominant \cite{Nojiri2004}. Moreover,  the evolution becomes
non-unitary in the achronal region around the big rip because of
the presence of wormholes whose creation is induced  by the exotic
character of the phantom energy \cite{Sushkov2005, Lobo2005, PFGD2003}. Therefore, a
generalized quantum theory  \cite{PFGD2009} has to be used to give
a proper quantum description of the whole phantom universe.

\section{Decoherence in a third quantization scheme}

\subsection{Parent and baby universes}

In a third quantization formalism \cite{Strominger1990}, the field to be quantized is the wave function of the universe. Then, the state of the multiverse can be studied as a quantum field theory in the superspace.

Let us consider the Wheeler-De Witt equation (\ref{WDW1}) with no
scalar field and without any factor ordering terms, i.e.
\begin{equation}\label{WDW}
\ddot{\phi} + \Omega^2(a) \phi = 0 ,
\end{equation}
where the overhead dot means derivative with respect to the scale
factor. Eq. (\ref{WDW}) can be seen as the classical equation of
motion for a harmonic oscillator with a \emph{time} dependent
frequency, with the scale factor playing the role of the time
variable. The wave function of the multiverse satisfies then the
Schr\"{o}dinger equation,
\begin{equation}\label{Schrodinger third quantized}
\mathbf{\mathrm{H}} |\Psi \rangle = i \hbar \frac{\partial}{\partial a} |\Psi \rangle ,
\end{equation}
with,
\begin{equation}\label{Hamiltonian 3-quantized}
\mathbf{\mathrm{H}} = \frac{1}{2} \hat{P}_\phi^2 + \frac{\Omega^2(a)}{2} \hat{\phi}^2 .
\end{equation}
$\hat{\phi}$ and $\hat{P}_\phi$ respectively are the operators of
the wave function of a single universe and its conjugate momentum,
in the Schr\"{o}dinger picture. Going into Heisenberg picture,
these operators can be written as
\begin{eqnarray}\label{H1}
\hat{\phi}(a) &=& A(a,a_0) \hat{\phi} + B(a,a_0) \hat{P}_\phi , \\ \label{H2}
\hat{P}_\phi(a) &=& \dot{A}(a,a_0) \hat{\phi} + \dot{B}(a,a_0) \hat{P}_\phi ,
\end{eqnarray}
where the functions $A(a,a_0)$ and $B(a,a_0)$ satisfy Eq. (\ref{WDW}) with
the initial conditions, $A(a_0,a_0) = \dot{B}(a_0,a_0) = 1$ and
$\dot{A}(a_0,a_0) = B(a_0,a_0) = 0$. The boundary condition that we impose
to the quantum state of the multiverse is that the number of
universes in the multiverse is constant along the evolution of the
scale factor within a single universe. Then, the state of the
multiverse is given in terms of the Lewis states \cite{Lewis1969,
RP2010}, which are defined by the following creation and
annihilation operators for universes,
\begin{eqnarray}\label{creationAnnihilation1}
b(a) & \equiv & \sqrt{\frac{1}{2 \hbar}} \left(\frac{\hat{\phi}}{R} + i (R   \hat{P}_\phi - \dot{R} \hat{\phi} ) \right) , \\  \label{creationAnnihilation2}
b^\dag(a) & \equiv & \sqrt{\frac{1}{2 \hbar}} \left(\frac{\hat{\phi}}{R} - i (R  \hat{P}_\phi - \dot{R} \hat{\phi} ) \right) ,
\end{eqnarray}
where $R\equiv R(a)$ is a function that satisfies the auxiliary equation, $\ddot{R} + \Omega^2(a) R - \frac{1}{R^3} = 0$. In terms of the operators (\ref{creationAnnihilation1}-\ref{creationAnnihilation2}), the third quantized Hamiltonian (\ref{Hamiltonian 3-quantized}) turns out to be
\begin{equation}\label{Hamiltonian2}
\mathbf{\mathrm{H}}  = \hbar \left[ \beta_- b^2 + \beta_+{b^\dag}^2 + \beta_0 \left( b^\dag b + \frac{1}{2} \right) \right] ,
\end{equation}
where \cite{RP2010},
\begin{eqnarray}\label{beta+}
\beta_+^* = \beta_- &=& \frac{1}{4} \left\{ \left( \dot{R} - \frac{i}{R} \right)^2 + \Omega^2 R^2  \right\} , \\ \label{beta0}
\beta_0 &=& \frac{1}{2} \left( \dot{R}^2 + \frac{1}{R^2} + \Omega^2 R^2 \right) .
\end{eqnarray}
We can consider large universes with a characteristic length of
order of the Hubble length of our universe. They will be called
parent universes  \cite{Strominger1990}. For large values of the
scale factor,  the non-diagonal terms in the Hamiltonian
(\ref{Hamiltonian2}) vanish and the values of the coefficient
$\beta_0$ asymptotically coincide with that of the proper
frequency of the Hamiltonian \cite{RP2010}. Then, the quantum
correlations between the number states disappear and the quantum
transitions between different number of universes are therefore
forbidden for parent universes. Let us also notice that in that
limit  the adiabatic approximation is satisfied,
$\frac{\dot{\Omega}}{\Omega} \ll \Omega$, and no creation of
further universes can occur along the evolution of a parent
universe.

Let us consider next the quantum fluctuations of the space-time of
a parent universe, whose contribution to the wave function of the
universe becomes important at the Planck scale \cite{Wheeler1957}.
Some of these fluctuations can be viewed as tiny regions of the
space-time that branch off from the parent universe and rejoin the
large regions thereafter; thus, they can be then interpreted as
virtual baby universes \cite{Strominger1990}. In that case
\cite{RP2010}, $\beta_+^* = \beta_-  \rightarrow
-\frac{\omega_0}{4}$ and $\beta_0 \rightarrow \frac{\omega_0}{2}$,
in Eq. (\ref{Hamiltonian2}), where $\omega_0$ is a constant that
depends on the properties of the baby universe. The state of the
gravitational vacuum is then represented by a squeezed state, with
a \emph{particle} creation of baby universes or fluctuations
occuring along the expansion of the parent universe
\cite{Grishchuk1990}.

Thus, parent and baby universes, which will be considered the
subsequent sections, can be described in the context of a third
quantization formalism as the states of a harmonic oscillator. The
advantage of such a formalism becomes then clear: we can apply the
well-studied machinery of harmonic oscillators and quantum field
theory for the description of a parent universe or a plasma of
baby universes. For instance, the propagator for the quantum state
of a parent universe can be calculated from the propagator for the
harmonic oscillator with time dependent frequency \cite{Khandekar1986} (see also Refs. \cite{Mukhanov2007,
Vergel2009}),
\begin{widetext}
\begin{eqnarray}\label{prop}
K(\Phi_f, a ; \Phi_i, a_0) &=& \left[ \frac{1}{2 \pi i \hbar R(a)
R(a_0) \sin\alpha(a,a_0)} \right]^{\frac{1}{2}} \exp\left[
\frac{i}{2 \hbar} \left\{ \frac{\dot{R}(a)}{R(a)} \Phi_f^2 -
\frac{\dot{R}(a_0)}{R(a_0)} \Phi_i^2 \right\} \right] · \\ & &
\exp\left[ \frac{1}{2 \hbar \sin\alpha(a,a_0)} \left\{ \left(
\frac{1}{R^2(a)} \Phi_f^2 + \frac{1}{R^2(a_0)} \Phi_i^2 \right)
\cos\alpha(a,a_0) - \frac{2}{R(a) R(a_0)} \Phi_i \Phi_f\right\}
\right],
\end{eqnarray}
\end{widetext}
where, $\Phi_f$ and $\Phi_i$, are the wave functions of the parent
universe evaluated at two hypersurfaces given by the values of the
scale factor $a$ and $a_0$, respectively, and $\alpha(a,a_0)$ is
defined as \cite{Lewis1969, RP2010}
$$
\alpha(a,a_0) = \int_{a_0}^a\frac{da'}{R^2(a')} .
$$
Another example is the density matrix that describes a heat bath
of  baby universes at temperature $T$. It can be written as the
density matrix of a canonical ensemble of harmonic oscillators
(see, for instance, Ref. \cite{Joos2003}),
\begin{widetext}
\begin{equation}
\rho_B(\phi_f, \phi_i) = \prod_n \frac{\omega_n}{2 \pi \sinh (\omega_n/kT)} \; \exp \left( -\frac{\omega_n}{2  \sinh (\omega_n/kT)} [ (\phi_{f,n}^2 + \phi_{i,n}^2) \cosh(\omega_n/kT) - 2 \phi_{f,n} \phi_{i,n} ]  \right) ,
\end{equation}
\end{widetext}
where, $\phi_{f,n}$ and $\phi_{i,n}$, are the wave functions of
the baby universes, which are represented by harmonic oscillators
with frequency $\omega_n$, and the index $n$ labels the species of
baby universes considered in the space-time foam.

\subsection{Parent-baby interaction}

Let us now pose the interaction scheme between a parent universe
and its environment. First, we shall study the interaction between
a parent universe and the quantum fluctuations of its space-time,
being these represented by a plasma of baby universes. The
gravitational vacuum will be considered in two different states:
i) the state of a heat bath  with temperature $T$, and ii) a
squeezed vacuum state. Then, it will be analyzed the interaction of a parent universe
with the rest of parent universes in the context of a multiverse.

The interaction between a parent universe and the quantum fluctuations of its space-time can be represented, in a first approximation, by a total Hamiltonian given by
\begin{equation}\label{totalH}
H = H_P + H_\varepsilon + H_{int} ,
\end{equation}
where $H_P$ is the Hamiltonian of the parent universe,
$H_\varepsilon$ is the Hamiltonian of the plasma of baby
universes, and $H_{int}$ is the interaction Hamiltonian. The
former Hamiltonian is represented by a harmonic oscillator with a
frequency that depends on the scale factor,
\begin{equation}
H_P = \frac{1}{2} P_\Phi^2 + \frac{\Omega^2(a)}{2} \Phi^2 ,
\end{equation}
with,  $q = \frac{3}{2}(1-w)$, and $\Omega(a) = \frac{\Omega_0}{q} a^{q-1}$, where $\Omega_0^2$ is proportional to the current energy density of our universe.

For the case of baby universes, the frequency of the harmonic
oscillator  can  effectively be considered a constant   determined
by the energy and the characteristic length of the baby universe,
which can go from the Planck length to the scale of laboratory
physics in the dilute-gas approximation \cite{Farhi1987, Kiefer2007,
Coleman1988, Coleman1988b}. It is therefore very small compared
with the large value of the length of the parent universe,  which
is of order of the Hubble length of our universe. Thus,  the
plasma of baby universes is represented in our model by a set of
harmonic oscillators with constant frequency $\omega_i$, where the
index $i$ labels the different species of baby universes, i.e.
\begin{equation}\label{Hbaby}
H_\varepsilon = \sum_{i=1}^N \frac{1}{2} p_{\phi_i}^2 + \frac{\omega_i^2}{2} \phi_i^2 ,
\end{equation}
The interaction Hamiltonian, $H_{int}$ in Eq. (\ref{totalH}), can be written as
\begin{equation}
H_{int} = \sum_i \lambda_i  \Phi_P \otimes f(\phi_i) ,
\end{equation}
where $\lambda_i$ is an effective coupling constant between the
parent universe and the baby universe  $i$, which is assumed to be
small so that the Born-Markov approximation can be assumed to hold
in the interaction scheme. The  form of $f(\phi_i)$ depends on the
kind of interaction which is considered to take place between the
parent and the baby universes. For instance, in the case
considered by Coleman \cite{Coleman1988, Coleman1988b} and others
\cite{Giddings1988}, in which simply-connected wormholes are
considered and thus single baby universes are nucleated,
$f(\phi_i) \equiv \phi_i$. In the case considered by
Gonz\'{a}lez-D\'{i}az \cite{PFGD1992a, PFGD1992b}, where
doubled-connected wormholes are created and therefore the baby
universes are nucleated in pairs, $f(\phi_i)\equiv \phi_i^2$. We
shall consider these two cases in the analysis to follow. In
general, $f(\phi_i)$ can be a complicated function  and these two
extreme cases can be considered as the first terms of its series
development.

As a result of the interaction between the parent universe and the
plasma of baby universes, the properties of the parent universe
are modified so that its evolution effectively becomes
non-unitary. The master equation for the reduced density matrix of
the parent universe, when the degrees of freedom of the baby
universes are traced out, can be written as
\cite{Schlosshauer2007, Joos2003}
 \begin{widetext}
\begin{equation}\label{eq5356}
\partial_a \rho_P = - i [\tilde{H}_P, \rho_P] - i \gamma [\Phi, \{ P_\Phi , \rho_P \}] - D [\Phi, [\Phi, \rho_P] ] - f [\Phi, [P_\Phi, \rho_P]] .
\end{equation}
\end{widetext}
The unitary part of the effective evolution of the parent
universe, given by the first term in Eq. (\ref{eq5356}),
corresponds to the evolution of a new harmonic oscillator,
$\tilde{H}_P \equiv H_P + \frac{\tilde{\Omega}^2}{2} \Phi^2$, with
a frequency which is shifted with respect to the initial value
$\Omega(a)$. The corresponding Lamb shift is given by
\begin{equation}\label{shift1}
\tilde{\Omega}^2(a) = -2  \int_{a_0}^a da' \, \eta(a, a') A(a, a') ,
\end{equation}
where $\eta(a, a')$ is the imaginary part of the correlation function,
\begin{equation}\label{correlationFunction}
\langle \phi^k(a) \phi^k(a')\rangle_b = \nu(a, a') - i \eta(a, a') ,
\end{equation}
with, $k=1$ and $k=2$, for  linear and  quadratic interactions,
respectively, and $A(a,a_0)$ is a solution of the Weeler-De Witt
equation (\ref{WDW}) (see,  Eqs. (\ref{H1}-\ref{H2})). The shift
for the frequency of the harmonic oscillator that represents the
state of the parent universe corresponds to a shift of its energy
density. Therefore, the Lamb shift given by Eq. (\ref{shift1}) can
be viewed to be equivalent to the mechanism proposed by Coleman to
set zero the most probable value of the cosmological constant into
zero. It is currently known that the value of the cosmological
constant is not zero but very small, in fact of the order of the
current critical density.

Three terms can be distinguished in the non-unitary part of the
master equation (\ref{eq5356}). The dissipation coefficient
\cite{Schlosshauer2007}, $\gamma(a)$, is given by
\begin{equation}
\gamma(a) =   \int_{a_0}^a da' \, \eta(a, a') B(a, a') ,
\end{equation}
where $B(a,a_0)$ is defined by Eqs. (\ref{H1}-\ref{H2}). In quantum
mechanics, $\gamma$ is related to the momentum damping and to the
velocity of the wave packet. However, the wave function of the
universe is not defined upon the space-time but in the superspace
so that the interpretation of $\gamma$ does not become so clear
for the state of the multiverse.

The normal-diffusion coefficient \cite{Schlosshauer2007}, $D(a)$ in Eq. (\ref{eq5356}), is given by
\begin{equation}
D(a) =  \int_{a_0}^a da' \, \nu(a, a') A(a, a') ,
\end{equation}
where $\nu(a, a')$ is the real part of the kernel (\ref{correlationFunction}). In quantum mechanics, $D$ gives a measure of the decoherence length of a Gaussian wave packet \cite{Schlosshauer2007}. In the case of the universe, this coefficient provides us therefore with a measure of the effectiveness of the decoherence process of two different branches of the universe, $\Phi$ and $\Phi'$,  caused by the interaction with the quantum fluctuations of the gravitational vacuum. Finally, the anomalous-diffusion coefficient \cite{Schlosshauer2007} in Eq. (\ref{eq5356}) is given by,
\begin{equation}\label{gamma1}
f(a) = -  \int_{a_0}^a da' \, \nu(a, a') B(a, a') .
\end{equation}

Let us now derive the results of in the third quantization
formalism for linear and quadratic interactions. Two states are
considered for the gravitational vacuum: i) a thermal state and,
ii) a squeezed vacuum state.

\subsubsection{Case 1: linear interaction}

The wave functions that represent the baby universes are given by
the solutions corresponding to the harmonic oscillator with
constant frequency. These can be written, in terms of the creation
and annihilation operators of baby universes, $b_i^\dag$ and
$b_i$, as
\begin{equation}\label{phibaby}
\phi_i(a) = \sqrt{\frac{1}{2 \omega_i}} \left( e^{- i \omega_i (a-a_0)} b_i + e^{ i \omega_i (a-a_0)} b_i^\dag \right) ,
\end{equation}
where $a$ is the scale factor of the parent universe, and $b_i^\dag$ and $b_i$, are the creation and annihilation of baby universes evaluated at the hypersurface given by  the value $a_0$. Then, for the case of linear interaction, $k=1$ in the correlation function (\ref{correlationFunction}). In the case of a thermal bath of baby universes, $\langle b_i^2 \rangle_b = \langle (b_i^\dag)^2 \rangle_b = 0$ and $\langle b_i^\dag b_i \rangle_b = N_i(T) = \langle b_i b_i^\dag \rangle_b -1$, with
\begin{equation}\label{N}
N_i(T) = \frac{1}{e^{\frac{ \omega_i}{ T}} - 1} ,
\end{equation}
where $T$ is the temperature of the space-time foam \cite{Garay1998}, in units $\frac{k_B}{\hbar}=1$. The noise kernel \cite{Schlosshauer2007}, $\nu(a,a')$, and the dissipation kernel \cite{Schlosshauer2007}, $\eta(a, a')$, in Eq. (\ref{correlationFunction}), can be written then as
\begin{eqnarray}\nonumber
\nu(a-a') &=& \sum_i \frac{ \lambda_i^2}{2 \omega_i} (2 N_i + 1) \cos \omega_i(a-a') , \\ \label{noise}
 &=& \int_0^\infty d\omega J(\omega) \coth\left( \frac{ \omega}{2  T} \right) \cos \omega(a-a')  \\  \nonumber
\eta(a-a') &=& \sum_i \frac{ \lambda_i^2}{2 \omega_i} \sin \omega_i(a-a') , \\ \label{dissipation}
 &=& \int_0^\infty d\omega J(\omega)  \sin \omega(a-a') ,
\end{eqnarray}
where, $J(\omega) \equiv \sum_i \frac{ \lambda_i^2}{2
\omega_i} \delta(\omega-\omega_i)$, is the spectral density of
baby universes in the space-time foam. It encapsulates the
physical properties of the plasma of baby universes. For the
quantum fluctuations of the gravitational vacuum, it is expected
that the presence of baby universes in the space-time foam be
exponentially suppressed for large values of  the energy of the
baby universe. Therefore, we assume the following spectral density
for the bath of baby universes, $J(\omega) = J_0^2 \omega^3
e^{-\frac{\omega}{\Lambda}}$, where $J_0$ and $\Lambda$ are two
constants, the latter representing the cut-off for the energy of
the vacuum fluctuations, and the factor $\omega^3$ has been
introduced to make the value of $J(\omega)$ sufficiently
convergent at $\omega\rightarrow 0$.

The functions $A(a,a_0)$ and $B(a,a_0)$ in Eqs. (\ref{H1}-\ref{H2}) and Eqs. (\ref{shift1}-\ref{gamma1}) can be expressed in terms of Bessel functions. For large values of the scale factor, which correspond to the description of parent universes, they can be approximated as
\begin{eqnarray}\label{eq5339}
A(a,a') &=& \left( \frac{a'}{a}\right)^{\frac{q-1}{2}} \cos\left(\frac{\Omega_0}{q}(a^q-a'^q)\right), \\ \label{eq5340}
B(a,a') &=&  ( a a' )^{\frac{1-q}{2}} \frac{1}{\Omega_0} \sin\left(\frac{\Omega_0}{q}(a^q-a'^q)\right) .
\end{eqnarray}
where $q=\frac{3}{2}(1-w)$. For, $q=1$ ($w=\frac{1}{3}$), these equations are exact and correspond to the solutions of a harmonic oscillator with constant frequency, $\Omega_0$. If we consider  small changes in the scale factor of the parent universe, $a - a_0 \ll 1$ and $a-a' \ll 1$, then,
\begin{eqnarray}\label{eq53}
\tilde{\Omega}^2(a) &\approx& - \frac{2 c_1 a^2}{n_1 n_2} \left( 1 - n_2 \left( \frac{a_0}{a} \right)^{n_1} + n_1\left( \frac{a_0}{a} \right)^{n_2}  \right) \\ \label{eq54} D(a) &\approx& \frac{c_2 a}{n_1} \left( 1 - \left( \frac{a_0}{a} \right)^{n_1} \right) ,
\end{eqnarray}
with, $n_1=\frac{q+1}{2}$, $n_2= \frac{q+3}{2}$, and
\begin{eqnarray}
c_1 &=& \int_0^\infty d\omega \, \omega J(\omega)  \approx  24 J_0^2 \Lambda^5 , \\ \label{c2}
c_2 &=& \int_0^\infty d\omega J(\omega)  \coth\left( \frac{\omega}{2  T} \right) \approx  4 J_0^2 \Lambda^3 T ,
\end{eqnarray}
where the limit $\frac{\omega}{T}\ll 1$, has been taken in the
latter equation.

On the other hand, if we describe the plasma of baby universes by a squeezed state, which can be considered a more realistic case \cite{Grishchuk1990, Kiefer2007}, then
\begin{eqnarray}
\langle b_i^\dag b_i \rangle &=& \sinh^2 r_i \equiv \tilde{N}_i, \\
\langle b_i b_i^\dag \rangle &=&  \cosh^2 r_i \equiv \tilde{N}_i + 1 , \\
\langle b_i^2 \rangle &=& -  \frac{1}{2} e^{i\theta_i}  \sinh 2r_i , \\
\langle (b_i^\dag)^2 \rangle &=& -  \frac{1}{2} e^{-i\theta_i} \sinh 2r_i ,
\end{eqnarray}
where $r_i$ and $\theta_i$ are the squeezing parameters. The two
first terms are equivalent to the case of a thermal bath of baby
universes with an effective number of quanta given by,
$\tilde{N}_i \equiv \sinh^2 r_i$. Thus, the dissipation kernel,
$\eta(a,a')$, turns out to be the same as in the thermal case, as
it is also given by Eq. (\ref{dissipation}). Then, the Lamb shift
$\tilde{\Omega}(a)$ is that given by Eq. (\ref{eq53}). However, the squeezed
vacuum introduces new terms in the noise kernel, $\nu(a,a')$. This
is given in the case of the squeezed vacuum by
\begin{eqnarray}\nonumber
\nu_s(a-a') &=& \sum_i \frac{\hbar \lambda_i^2}{2 \omega_i} \{ (2 \tilde{N}_i + 1) \cos \omega_i(a-a') \\ &-& \sinh 2r_i \cos(\omega_i(a+a'-2a_0) -\theta_i) \} .
\end{eqnarray}
Then, the decoherence factor, $D(a)$, turns out to be given by Eq. (\ref{eq54}), with
\begin{equation}\label{c22}
c_2^s =  \int_0^\infty d\omega J(\omega) \left( \cosh 2r - \sinh 2r \cos\theta  \right) .
\end{equation}
A similar expression is obtained in Ref. \cite{Joos2003}, p. 217
(see, also, Ref. \cite{Kiefer1998}). There, a decoherence
timescale is given by, $t_D^{-1} = \sqrt{c_2^s}$. In the limit of
large squeezing \cite{Joos2003}, $\theta \rightarrow 0$ and $r
\rightarrow -\infty$, and we can estimate a decoherence scale for
two different branches of the parent universe given by, $a_D
\approx \frac{1}{J_0 \Lambda^2 e^{|r|}}$.

In both, a vacuum in a thermal and in a squeezed state, the effect
of decoherence due to the interaction of the parent universe with
the quantum fluctuations of the space-time is similar if
effectively we assume that, $T \sim \Lambda N$
in the thermal bath, and that, $e^{2|r|}\sim N$, in the squeezed
vacuum; that is, for a large number of fluctuations of the
space-time.

On the other hand, the timescale for the decoherence of a wave
packet is differently analyzed in Ref. \cite{Schlosshauer2007}.
There, the decoherence factor  $D$ measures the decoherence of a
Gaussian wave packet at spatial positions $x$ and $x'$, with a
decoherence time given by \cite{Schlosshauer2007}, $\tau_D =
\frac{1}{D (x -x')^2}$. In the case of the universe, $\Phi$ and
$\Phi'$ represent different branches of the parent universe, and
therefore, a decoherence scale of order $a_D \sim \frac{1}{c_2
(\Phi -\Phi')^2}$ can be assumed, with $c_2$ as given by Eq.
(\ref{c2}) or Eq. (\ref{c22}) for the case of a thermal
bath or a squeezed vacuum, respectively.

In any case, it can be concluded that the scale at which quantum
interference between different branches of a parent universe
becomes important is very small, presumably of order the Planck
length.

\subsubsection{Case 2: quadratic interaction}

In the case of a quadratic interaction between the parent universe
and the baby universes, $k=2$ in the correlation function given by
Eq. (\ref{correlationFunction}). The formalism applies in the same
way as in all the previous cases and the quadratic interaction
only changes the functional form of the dissipation and noise
kernels, $\eta(a,a')$ and $\nu(a,a')$, respectively. For a vacuum
in a thermal state and $a-a_0 \ll 1$, they are given by
\begin{eqnarray}
\eta(a,a') &=& 4  \int_0^\infty d\omega  J(\omega) N (a-a'), \\
\nu(a,a') &=& 2  \int_0^\infty d\omega  \frac{J(\omega)}{\omega} N^2 ,
\end{eqnarray}
where, $N\equiv N(\omega)$, is defined in Eq. (\ref{N}). Then, the Lamb shift, $\tilde{\Omega}(a)$, and the decoherence factor, $D(a)$, are those given by Eqs. (\ref{eq53}) and (\ref{eq54}), respectively, with new coefficients $d_1$ and $d_2$ instead of $c_1$ and $c_2$, given by
\begin{eqnarray}
d_1 &=& 8 T \int_0^\infty d\omega \; \frac{J(\omega)}{\omega} \approx 16 J_0^2 \Lambda^3 T , \\ \label{d2}
d_2 &=& 8 T^2 \int_0^\infty d\omega \frac{J(\omega)}{\omega^3} \approx 8 J_0^2 \Lambda T^2 .
\end{eqnarray}
For a squeezed vacuum state, the leading terms of the dissipation and noise kernels turn out to be
\begin{eqnarray}
\eta(a,a') &\approx&  \int_0^\infty d\omega J(\omega) e^{2|r|} (a-a') , \\
\nu(a,a') &\approx&  \int_0^\infty \frac{J(\omega)}{2 \omega} e^{4 |r|}  ,
\end{eqnarray}
and then
\begin{eqnarray}
d_1^s & \approx & 6 J_0^2 \Lambda^4 e^{2|r|} , \\ \label{d22}
d_2^s & \approx & J_0^2 \Lambda^3 e^{4|r|} .
\end{eqnarray}
For the quadratic interaction, therefore, the coefficient $d_1$
that determines the Lamb shift depends on the temperature, for  a
thermal vacuum, and on the squeezing parameter $r$, for a squeezed
vacuum state. It depends thus on the strength of the fluctuations
of the space-time of the parent universe, which is assumed to be
large. The coefficient of the decoherence factor, $d_2$, depends
on $\tilde{N}_i^2$ in the quadratic interaction. However, this
kind of interaction is of order $\hbar^2$ instead of $\hbar$ for
linear interaction, and therefore the contribution of the
quadratic interaction is subdominant in the semiclassical regime
of the quantum state of the universe.

\subsection{Parent-parent interaction}

We can also consider the interaction between a parent universe and
the rest of universes of a multiverse made up of parent universes.
In that case, the squeezing effect of the state of the multiverse
asymptotically disappears \cite{RP2010}, i.e. $r\rightarrow 0$ as
$a\rightarrow \infty$. It seems then most appropriate considering
a thermal state of $N$ parent universes, with
$N^{-1}=e^{\frac{\omega}{T}}-1$, where $T$ is a temperature analog
in the multiverse. It includes the special case for which $N=0$
($T=0$) that represents the interaction of a parent universe with
the fluctuations of its ground state.

For parent universes, the coefficients of their wave functions can be approximated by Eqs. (\ref{eq5339}-\ref{eq5340}), and then
\begin{widetext}
\begin{eqnarray}
\tilde{\Omega}^2(a) & \approx & -\frac{2}{a^{q-1}} \int_{a_0}^a \int_0^\infty d\Omega' J(\Omega') \sin\frac{\Omega'}{q}(a^q-a'^q) \cos\frac{\Omega_0}{q}(a^q-a'^q) , \\
D(a) & \approx & \frac{1}{a^{q-1}} \int_{a_0}^a \int_0^\infty d\Omega' J(\Omega') (2 N+1) a'^{q-1} \cos\frac{\Omega'}{q}(a^q-a'^q) \cos\frac{\Omega_0}{q}(a^q-a'^q) ,
\end{eqnarray}
\end{widetext}
where  $J(\Omega')$ is now the spectral density of parent
universes in the multiverse, and $\Omega'$ refers to the energy
density of parent universes, which is assumed to be picked around
the current energy density of our universe. For small intervals of
the scale factor,
\begin{eqnarray}
\tilde{\Omega}^2(a) & \approx & -\frac{2 c_1^p a^2}{qn} \left( q +\left( \frac{a_0}{a}\right)^{n} -n \left( \frac{a_0}{a}\right) \right)  \\
D(a) & \approx & \frac{c_2^p a}{q} \left(1-\left( \frac{a_0}{a}\right)^q\right) ,
\end{eqnarray}
where $n=q+1$, and
\begin{eqnarray}
c_1^p &=& \int_0^\infty d\Omega' J(\Omega') \Omega' , \\ \label{c2p}
c_2^p &=& \int_0^\infty d\Omega' J(\Omega') \coth\frac{\Omega'}{2 T} .
\end{eqnarray}
If we assume that the energy density of the parent universes of
the multiverse is highly peaked around the value of the
theoretical value of the energy density of our universe,
$\Omega_0$, then $J(\omega')\sim \delta(\Omega'-\Omega_0)$, $c_1^p
\sim \Omega_0$, and $c_2^p \sim \coth\frac{\Omega_0}{2 T}$. The
corresponding Lamb shift given by $c_1^p$ is therefore of the same
order as the energy density that corresponded to a universe
without any interactions. Then, the effective energy density of
the universe turns out to be approximately zero. Furthermore, with
the same choice of spectral density in the multiverse, the
decoherence between two different branches of a parent universe is
effective for a large number of universes in the environmental
multiverse, i.e., when $c_2^p \sim \frac{2 T}{\Omega_0}\gg 1$.
However, these results are highly dependent on the choice of the
spectral density of the multiverse.

For other values of the interval $a-a_0$ rather than $a-a_0 \ll
1$, numerical methods have to be employed and the results strongly
depend on the estimation of the relative value of the energies of
the universes, and on the choice taken for the spectral density. A
particular simple case is when $q=1$, i.e. for universes which are
dominated by a radiation-like fluid ($p=\frac{1}{3}\rho$). Then,
the quantum state that describes the universes is that of a
harmonic oscillator with constant frequency where the
approximations used in Eqs. (\ref{eq5339}-\ref{eq5340}) become
necessarily exact manipulations, with $q=1$. In such a case,
assuming a large interval of interaction ($a-a_0\rightarrow
\infty$) and a spectral density given by $J(\omega)=J_0^2 \omega^3
e^{-\frac{\omega}{\Lambda}}$, it is obtained
\begin{eqnarray}\nonumber
D &=& \int_0^\infty d\tau \int_0^\infty d\omega J(\omega) (2 N + 1) \cos\omega\tau \cos\Omega_0\tau \\
&=& \pi J_0^2 \, \Omega_0^2 \, T \, e^{-\frac{\Omega_0}{\Lambda}}
. \label {D1}
\end{eqnarray}
$\Lambda$ in Eq.(\ref{D1}) is a cut-off for the energy density of
the environment and $\Omega_0$ the energy density of a
distinguished universe which herewith refers to ours own. For an
environment of baby universes, $\frac{\Omega_0}{\Lambda}\gg 1$,
and the decoherence between two branches of the parent universe is
only effective for large number of vacuum fluctuations or,
equivalently, for large values of the squeezing parameter $r$. For
an environment made up of parent universes,
$\frac{\Omega_0}{\Lambda}\sim 1$, and the decoherence effect is
more effective even for the interaction of the parent universe
with the fluctuations of its ground state, for which $D\sim
\Omega_0^3$.

\begin{figure}[h]

\includegraphics[width=8cm]{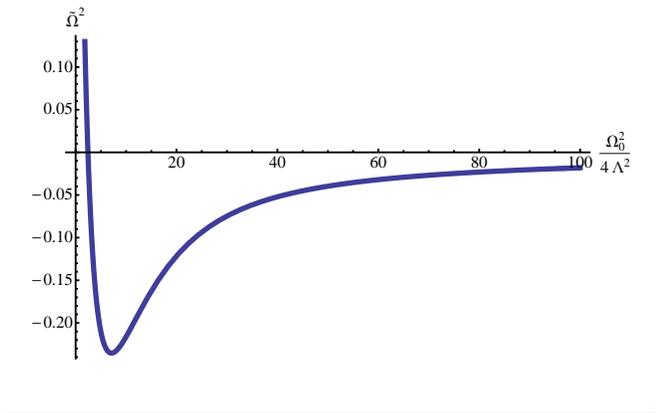}

\caption{The Lamb shift,  Eq. (\ref{Meijer}), in units $4\sqrt{\pi} J_0^2 \Lambda^3$.}.

\label{figura1}

\end{figure}

Moreover, assuming the same spectral density as for Eq.(\ref{D1})
and $a-a_0\rightarrow \infty$, the Lamb shift can be expressed in
terms of a Meijer G-function, i.e.
\begin{eqnarray}\nonumber
\tilde{\Omega}^2 &=& - 2 \int_0^\infty d\tau \int_0^\infty d\omega J(\omega)  \sin\omega\tau \cos\Omega_0\tau \\ \label{Meijer}
&=& 4\sqrt{\pi} J_0^2 \Lambda^3 {\rm G}_{1 3}^{2 1}\left( \begin{array}{lll} 0 \\ 0,3,\frac{1}{2}\end{array} \bigg| \frac{\Omega_0^2}{4 \Lambda^2}\right) ,
\end{eqnarray}
which is depicted in Fig. \ref{figura1}. For an environment made
up of baby universes, $\frac{\Omega_0}{\Lambda}\gg 1$ and  the
Lamb shift turns out to be very small. However, for an environment
of parent universes, $\Omega_0 \sim \Lambda$ and the corresponding
Lamb shift can be of order of the original frequency, resulting
then in an effective value of the energy density of the universe
very closed to zero.

The multiverse of parent universes turns out to be then more effective for both the decoherence between two branches and the reduction of the theoretical value of the vacuum energy density of our universe.

\subsection{Thermodynamical quantities}

\begin{figure}[h]

\includegraphics[width=8cm]{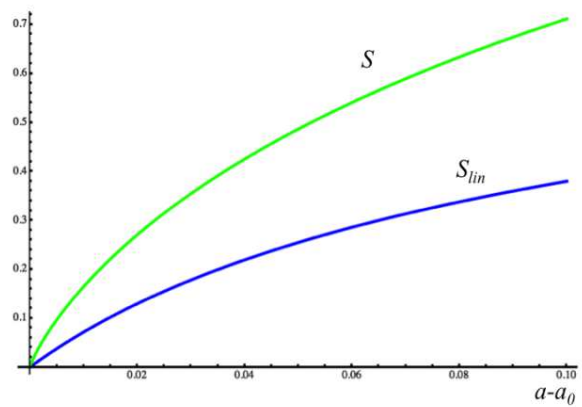}

\caption{Entropy and linear entropy of the state of a parent universe which is interacting with a plasma of baby universes, for a linear interaction (see, Eqs. (\ref{eq5382}-\ref{eq5383})).}

\label{fig:entropiaPureza}

\end{figure}

As a consequence of the interaction of a single universe with an
environment made up of baby or parent universes, the universe
undergoes an effectively non-unitary evolution determined by the
three last terms of the master equation (\ref{eq5356}). As a
simple example, let us take the value $q=1$, so that
$D\propto(a-a_0)$, $\gamma \propto(a-a_0)^3$, and
$f\propto(a-a_0)^2$. For a small value of the interval $\Delta
a\equiv a-a_0$, we can consider only the decoherence factor
$D(a)$. The master equation (\ref{eq5356}) can be written then, in
the configuration space, as
\begin{widetext}
\begin{equation}\label{eqMaster}
\partial_a \rho_P(\Phi, \Phi', a) =  \{ -\frac{i}{2} \left( \frac{\partial^2}{\partial \Phi'^2} - \frac{\partial^2}{\partial \Phi^2} \right) - \frac{i\Omega^2(a)}{2} (\Phi^2 - \Phi'^2)    - D(a) (\Phi - \Phi')^2   \} \rho_P(\Phi, \Phi', a) ,
\end{equation}
\end{widetext}
with, $D(a) \approx c_2 (a-a_0)$, where $c_2$ is given by Eqs. (\ref{c2},\ref{c22},\ref{d2},\ref{d22},\ref{c2p}) for the different kinds of interactions considered in this section, and $\Omega(a)\approx \Omega_{ef}(a_0)$. The master equation (\ref{eqMaster}) can be solved with the Gaussian ansatz \cite{Joos2003, Schlosshauer2007},
\begin{equation}\label{eq5373}
\rho(\Phi, \Phi', a) = e^{- A(a) (\Phi - \Phi')^2 - i B(a) (\Phi^2 - \Phi'^2) - C(a) (\Phi + \Phi')^2 - N(a) } .
\end{equation}
The coefficients $A(a)$, $B(a)$ and $C(a)$ satisfy then the following differential equations \cite{Joos2003} ($N$ is a normalization factor),
\begin{eqnarray}\label{eq5374}
\dot{A} &=& 4 A B + D(a) , \\ \label{eq5375} \dot{B} &=& 2 B^2 - 8 A C + \frac{1}{2} \Omega^2 , \\ \label{eq5376} \dot{C} &=& 4 B C .
\end{eqnarray}
In order to analyze the decoherence effects of the environment on
the universe, let us consider a separable initial state for the
distinguished  universe, i.e. a pure state, given by
\cite{Schlosshauer2007}
\begin{equation}\label{eq5377}
\rho(\Phi, \Phi', a_0) = (\frac{1}{2 \pi b^2})^{\frac{1}{2}} e^{- \frac{\Phi^2 + \Phi'^2}{4 b^2} } .
\end{equation}
In that case, the initial conditions for the coefficients $A(a)$,
$B(a)$ and $C(a)$ are $A(a_0) = C(a_0) = \frac{1}{8 b^2}$, and
$B(a_0) =0$. With the assumption, $ \Delta a \ll 1$, and
disregarding higher orders  than $\Delta a$, it is obtained (see,
App. A2 in Ref. \cite{Joos2003})
\begin{eqnarray}
A(a-a_0) &\approx& \frac{1}{8 } (1 + 16 C_0 (a-a_0)) , \\ B(a-a_0) &\approx&  \frac{C_0}{\Omega_0^2}  (a-a_0)  , \\ C(a-a_0) &\approx& \frac{1}{8}   ,
\end{eqnarray}
where, $C_0 = \Omega_0^4 - \frac{c_2}{8}$ (with, $b=1$). These
coefficients allow us to obtain the thermodynamical properties of
the parent universe. For instance, the purity of the state,
$\zeta$, is given by  \cite{Schlosshauer2007}
\begin{equation}
\zeta(a-a_0) = \sqrt{\frac{C(a)}{A(a)}} \approx \frac{1 }{\sqrt{1 + 16 C_0 (a-a_0)}} .
\end{equation}
The linear entropy \cite{Joos2003}, $S_{lin} \equiv {\rm Tr} (\rho-\rho^2)$, turns out to be
\begin{equation}\label{eq5382}
S_{lin} = 1 - \zeta \approx 1 - \frac{1 }{\sqrt{1 + 16 C_0 (a-a_0)}}  ,
\end{equation}
and the entropy of the distinguished universe, which for the
initial state is zero as corresponds to a pure state (see, Eq.
(\ref{eq5377})), grows up due to the interaction with the
environment according to
\begin{equation}\label{eq5383}
S = - \frac{1}{p_0} (p_0 \ln p_0 + q_0 \ln q_0 ) ,
\end{equation}
where \cite{Joos2003},
\begin{eqnarray}
p_0 &=& \frac{2 \zeta }{1 + \zeta} , \\ q_0 &=& \frac{1-\zeta}{1+\zeta} .
\end{eqnarray}
The linear entropy and the entropy given by Eqs. (\ref{eq5382})
and (\ref{eq5383}), respectively, are depicted in Fig.
\ref{fig:entropiaPureza} in units for which $C_0=1$, so that it is
qualitatively valid for all the kinds of interactions and
environments considered in this section. The interaction between
the parent universe and environment (made up of baby or parent
universes) makes the state of the universe to evolve into a mixed
state. That means that there exists a loss of information in the
state of the distinguished universe as a consequence of the
interaction with the quantum fluctuations of the space-time or
with other universes of the multiverse. That loss of information
makes the different branches of the universe to lose their quantum
coherence and,  together with other decoherence processes, leads
to the feature that the universe can be described in terms of the
semiclassical branch which we live in. It is worth noticing that
such a loss of information appears as a result of the trace
operation of the degrees of freedom that corresponds to the
environment. The total system, formed by the parent universe and
the rest of universes (baby or parent), retains all the
information of the system along the evolution of the multiverse.

\section{Conclusions}

The interaction between the scale factor and a scalar field with
mass is seen to decohere the expanding and contracting branches of
a geometrically flat, homogeneous and isotropic universe, whose
expansion (or contraction)  is accelerated. The decoherence turns
out to be more effective in the case of a universe dominated by a
quintessence fluid than when a vacuum or a phantom dominated
universe are considered. This might be related to the
crystal-clearer quantum nature of the latter universes.

The interaction of a parent universe with the environment, being
this formed by a multiverse of parent or baby universes, can be
analyzed following a parallel development to what is usually made
in quantum optics. Within the approximations considered in this
paper, the squeezed vacuum state and the thermal state of baby
universes produce similar effects provided that the squeezing of
the state of baby universes be interpreted as an effective
creation of a high number of fluctuations, i.e. for a large
squeezing effect.

The linear interaction produces leading terms in the change of the
properties of the parent universe, being therefore subdominant the
effects of the quadratic interaction. However, it does not imply
that the quadratic interaction has no relevant effects on the
state of the parent universe because it is actually the
responsible for the change of the high order coherence properties
of the fields that propagate upon the space-time \cite{PFGD1992b}.

The distinguished universe undergoes an effectively non-unitary
evolution as due to the interaction with the environment. The
decoherence and dissipation effects are even more acute if the
environment is taken to be a multiverse made up of parent
universes, because their energy density is assumed to be of the
order of that for the distinguished parent universe.

 Much as it happens with the Lamb shift in quantum mechanics, here the vacuum energy of the distinguished universe is also shifted. In the case of an environment made up of baby universes, which can be considered to be similar to that previously studied by Coleman \cite{Coleman1988}, the corresponding Lamb shift is important for a large squeezing effect or a large number of vacuum fluctuations. The effect is greater if the environment is a multiverse of parent universes since the corresponding Lamb shift matches the theoretical predictions for the vacuum energy of a single universe. That could effectively reduce the value of the energy density of the universe to be very closed to zero even for the interaction between the parent universe with the fluctuations of its ground state.

 Moreover, the entropy of the distinguished universe grows as a consequence of the interaction with its environment. Such an irreversible interaction may provide us then with an arrow of time.

\bibliographystyle{apsrev}

\bibliography{bibliography}

\end{document}